\documentclass[pra,nofootinbib,twocolumn,superscriptaddress,showpacs,showkeys]{revtex4-1}

\usepackage{graphicx,epsfig}
\usepackage{amsfonts,amsmath,amssymb,amsthm,amscd}
\usepackage{color}
\usepackage{dcolumn}
\usepackage{bm}
\usepackage{array}
\usepackage{float}
\usepackage[]{units}
\usepackage[colorlinks=true,linkcolor=blue,citecolor=blue]{hyperref}

\definecolor{teal}{rgb}{0, 0.5, 0.5}
\definecolor{darkmagenta}{rgb}{0.55, 0.0, 0.55}

\newcommand{\citeneeded}[1]{{\color{red}[citation needed]}}

\newcommand{\e}{\mathrm{e}}

\newcommand{\zR}{z_\mathrm{R}}
\newcommand{\Efront}{E_\mathrm{front}}
\newcommand{\Eback}{E_\mathrm{back}}

\begin{document}

\title{Effect of electron-beam energy chirp on signatures of radiation reaction\\ in laser-based experiments}

\author{J. Magnusson}
\affiliation{Department of Physics, University of Gothenburg, SE-41296 Gothenburg, Sweden}

\author{T. G. Blackburn}
\affiliation{Department of Physics, University of Gothenburg, SE-41296 Gothenburg, Sweden}

\author{E. Gerstmayr}
\affiliation{The John Adams Institute for Accelerator Science, Blackett Laboratory, Imperial College London, London SW7 2AZ, UK}
\affiliation{Centre for Light-Matter Interactions, School of Mathematics and Physics, Queen's University Belfast, Belfast BT7 1NN, United Kingdom}

\author{E. E. Los}
\affiliation{The John Adams Institute for Accelerator Science, Blackett Laboratory, Imperial College London, London SW7 2AZ, UK}

\author{M. Marklund}
\affiliation{Department of Physics, University of Gothenburg, SE-41296 Gothenburg, Sweden}

\author{C. P. Ridgers}
\affiliation{York Plasma Institute, School of Physics, Engineering and Technology, University of York, York, YO10 5DD, UK}

\author{S. P. D. Mangles}
\affiliation{The John Adams Institute for Accelerator Science, Blackett Laboratory, Imperial College London, London SW7 2AZ, UK}

\begin{abstract}
Current experiments investigating radiation reaction employ high energy electron beams together with tightly focused laser pulses in order to reach the quantum regime, as expressed through the quantum nonlinearity parameter $\chi$. Such experiments are often complicated by the large number of latent variables, including the precise structure of the electron bunch. Here we examine a correlation between the electron spatial and energy distributions, called an energy chirp, investigate its significance to the laser-electron beam interaction and show that the resulting effect cannot be trivially ignored when analysing current experiments. In particular, we show that the energy chirp has a large effect on the second moment of the electron energy, but a lesser impact on the first electron energy moment or the photon critical energy. These results show the importance of improved characterisation and control over electron bunch parameters on a shot-to-shot basis in such experiments.
\end{abstract}
\maketitle

\section{Introduction}
Recent advances in \emph{laser wakefield acceleration} (LWFA) have opened up the possibility for all-optical Compton scattering experiments for probing \emph{radiation reaction} (RR) in strong laser fields, now bordering the quantum regime~\cite{cole.prx.2018, poder.prx.2018}. This is made possible by the increased control of both the laser~\cite{danson.hplse.2019, kiriyama.hedp.2020, yoon.optica.2021} and the electron source~\cite{tajima.prl.1979, esarey.rmp.2009, kim.prl.2013, wang.ncomm.2013, leemans.prl.2014, gonsalves.prl.2019}, as well as improvements to their combination in collisional setups~\cite{bulanov.nima.2011, chen.prl.2013, sarri.prl.2014, yan.nphot.2017}.

The strong-field quantum regime is particularly enticing as it presents a largely unexplored domain~\cite{dipiazza.rmp.2012, gonoskov.rmp.2022, fedotov.pr.2023}, with several upcoming experimental campaigns aimed at exploring this regime~\cite{weber.mre.2017, gales.rpp.2018, abramowicz.arxiv.2019, meuren.arxiv.2020, salgado.njp.2022, turner.epjd.2022}. The quantum regime is defined through the \emph{quantum nonlinearity parameter} $\chi = |F_{\mu\nu}p^\nu|/mcE_\mathrm{S}$, with quantum behaviour emerging as $\chi$ approaches unity, where $F$ is the electromagnetic field tensor, $p$ is the electron four-momentum, $E_\mathrm{S}$ is the Sauter-Schwinger field strength~\cite{sauter.zp.1931, heisenberg.zp.1936, schwinger.pr.1951}, $m$ is the electron mass and $c$ the speed of light. Reaching the quantum regimes thus requires a combination of high field strength and high particle energy, and upcoming experiments are expected to rely on tight laser focusing in order to reach sufficiently high field intensities.

One of the first milestone in exploring the quantum regime experimentally is to measure the effect of radiation reaction to such accuracy that it becomes possible to discriminate between different RR models~\cite{cole.prx.2018,poder.prx.2018}. To do so requires good knowledge of the electron beam and laser pulse properties, but because these may fluctuate substantially between individual shots it is often not possible to evaluate the interaction on a shot-to-shot basis~\cite{samarin.jmo.2017}, necessitating a more statistical approach~\cite{maier.prx.2020, blackburn.rmpp.2020}.

There are several theoretical studies that investigate various aspects of electron-laser experiments directly related to radiation reaction, describing effects such as 
stochastic broadening~\cite{neitz.prl.2013, ridgers.jpp.2017}, straggling~\cite{shen.prl.1972, blackburn.prl.2014, harvey.pra.2016, geng.cp.2019}, and quenching~\cite{harvey.prl.2017}. A number of studies are aimed at increasing the amount of information gained per shot~\cite{streeter.hplse.2023}, \emph{e.g.} by utilising an astigmatic spot~\cite{Baird.njp.2019}, angular profiling of radiation emission~\cite{shemesh.ol.2012, olofsson.pra.2022} or finding an optimal parameter range~\cite{Arran.ppcf.2019}. Nevertheless, electron-laser experiments accommodate a large number of unknowns, many of which are often omitted in numerical and theoretical investigations in order to maintain a reasonable scope.

\begin{figure}[b!]
\includegraphics[width=0.85\columnwidth,trim={0cm 1.5cm 4cm 1cm},clip]{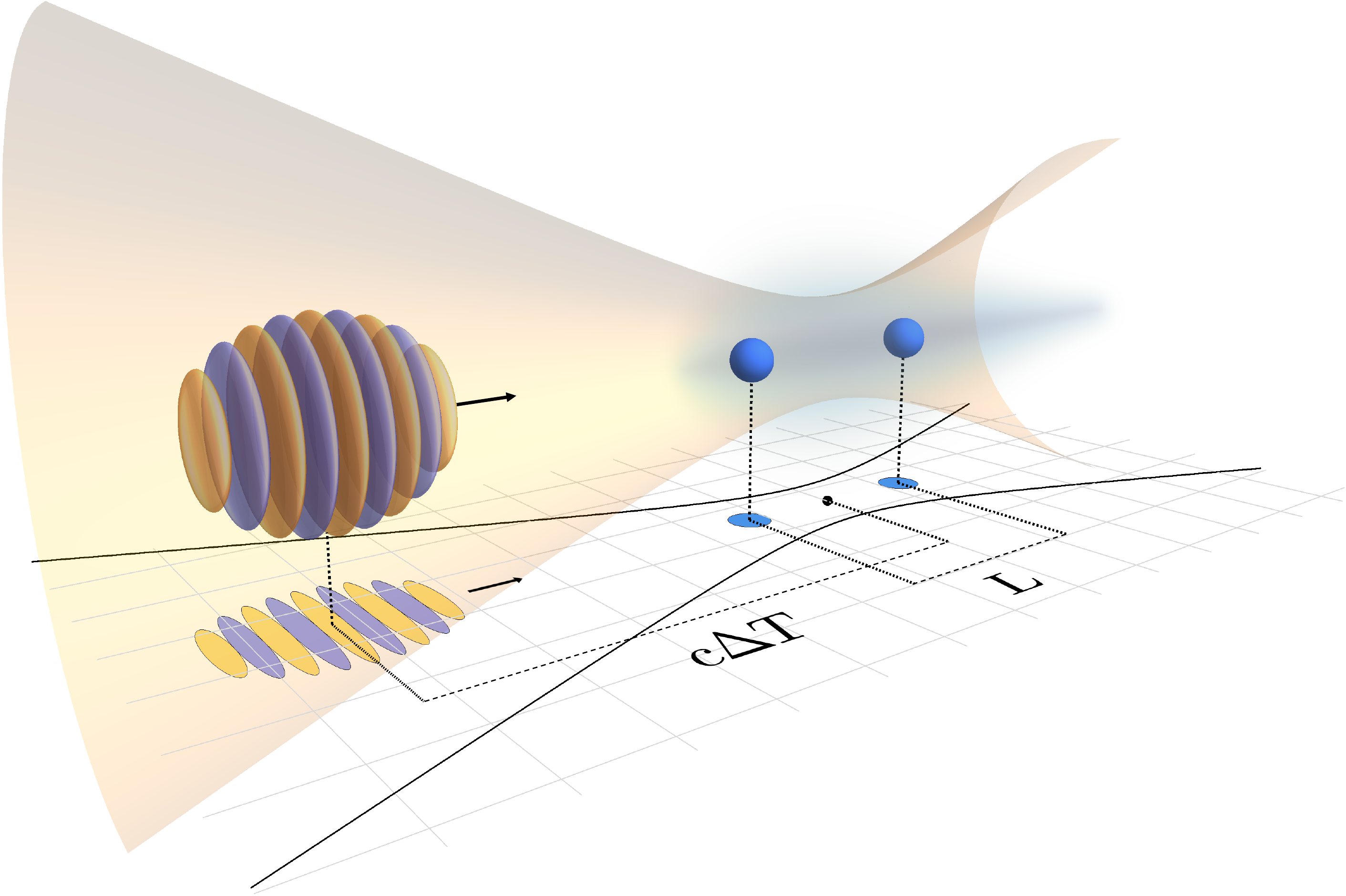}
\caption{\label{fig:setup} Schematic of a tightly focused laser pulse colliding with two electron micro-bunches, which are separated by a distance $L$, with a synchronisation offset of $c\Delta T$. The two bunches can have different energies and since they feel different field strengths, their relative position will affect the outcome of the interaction.}
\end{figure}

One such aspect relates to the phase space of the electron beam. While the energy spectrum of the electron beam is often examined, the same can not be said for the longitudinal density profile. Even then, any potential correlation between longitudinal position and energy has remained largely overlooked, and most studies model the electron beam as homogeneous~\cite{harvey.prab.2016}. For many cases this is perfectly valid, as most parts of the electron beam can often be expected to experience the same field strengths, albeit at different points in time. However, and as we shall show, this is no longer valid when when considering tightly focused laser fields, where the important parameter is the electron beam length $L$ relative to the \emph{Rayleigh range} $\zR$. Moreover, electron beams generated through LWFA are accelerated across an extended injection event, which often leads to a relatively broad energy spectrum that also contains a longitudinal chirp~\cite{dopp.prl.2018, brinkmann.prl.2017}.

The collision of such an electron bunch with a tightly focused laser pulse opens the possibility for different spectral ranges to experience different field strengths. The focus of this paper is to ascertain the size of this effect on the outcome of a laser-electron beam interaction and to determine the most important parameters for the effect. In section~\ref{sec:motivation} we analyse the problem analytically by studying the difference in field strength felt by two longitudinally displaced electrons, identifying the most important parameters. In section~\ref{sec:1d-sims} we perform a large number of single-particle simulations of monochromatic micro-bunches, focusing on mean electron energy and energy spread, and show the difference between the most common RR models. In section~\ref{sec:chirped-sims} we show how the energy chirp of a finite-sized electron beam affects the mean energy, standard deviation and photon critical energy after interacting with a tightly focused laser pulse.

\section{Motivation}\label{sec:motivation}
We begin by deriving a simple analytical estimate for when an energy chirp may become of importance. We do this by imagining two electrons co-propagating along the $z$-axis a distance $L$ apart, and counter-propagating to a tightly focused laser pulse, as shown in Figure~\ref{fig:setup}.

\begin{figure}[t!]
\includegraphics[width=1\columnwidth]{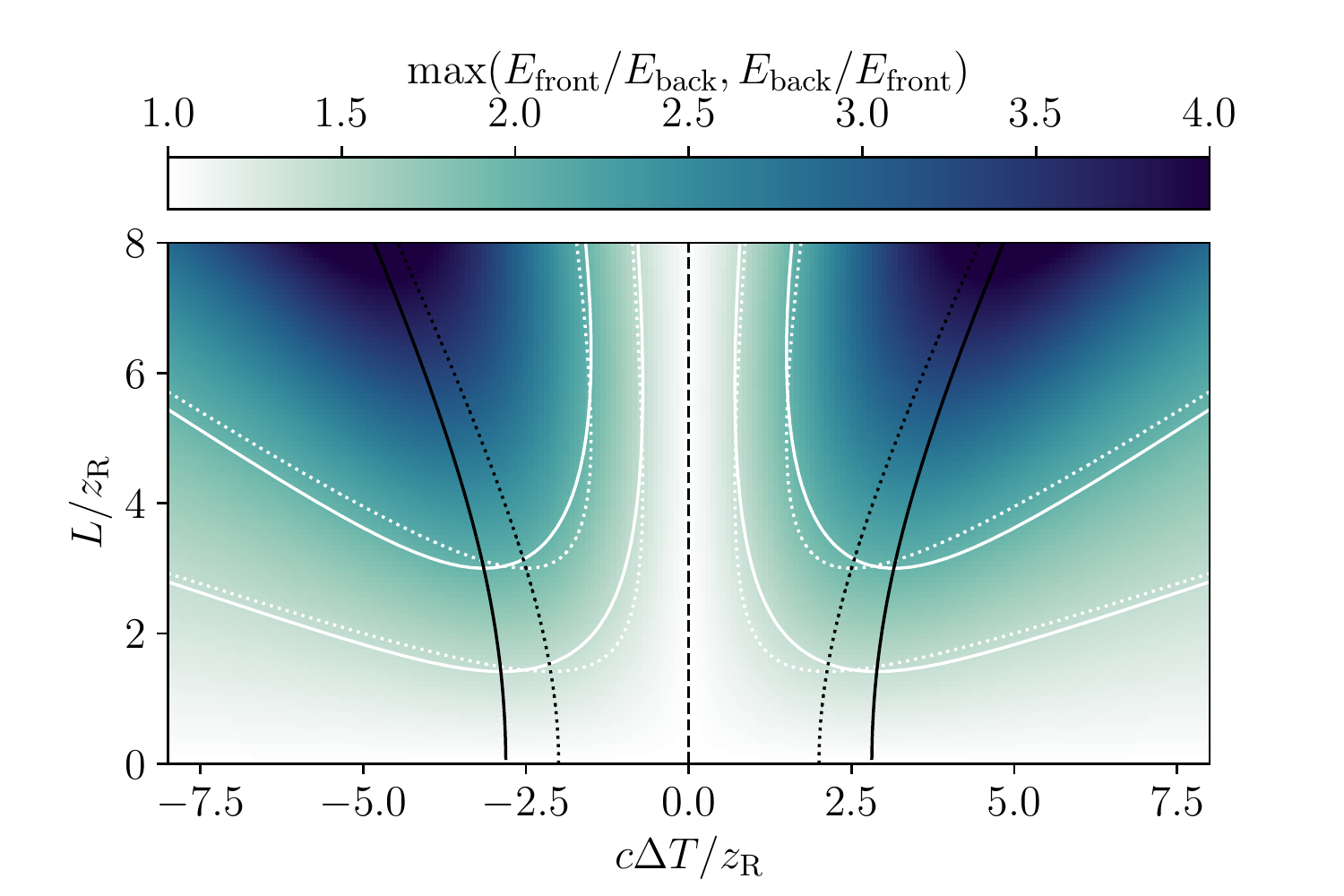}
\caption{\label{fig:amp}Ratio of the maximum field amplitude experienced by two electrons $L$ distance apart and with a synchronisation offset of $c\Delta T$ between the centre of mass of the two electrons and the laser pulse. The laser pulse duration is set to $c\tau = 3\zR$, which with $f/2$-focusing and $\lambda=\unit[0.8]{\mu m}$ is equivalent to $\unit[12]{\mu m}$ (\unit[40]{fs}) FWHM. Contours are presented for a ratio of $\sqrt{2}$ and $2$ (solid, white) as well as for the analytical approximation of equation~\ref{eq:amp} (dotted, white), the latter corresponding to $c\tau = 0$. The positions of the optima are also shown for each value of $L/\zR$ (solid, black) as well as for the analytical approximation of equation~\ref{eq:??} (dotted, black).}
\end{figure}

To first approximation, the field strength of a focused Gaussian laser pulse propagating in the negative $z$-direction can be obtained from the paraxial approximation as
\begin{equation}
E = E_0 \frac{w_0}{w(z)}\exp\!\left(-\frac{r^2}{w(z)^2}\right) \exp\!\left(-\left(\frac{kz+ct}{c\tau}\right)^2\right),
\end{equation}
where we have left out the phase factor and where
\begin{equation}
w(z)=w_0\sqrt{1+\left(z/\zR\right)^2}, 
\end{equation}
and $\zR = \pi w_0^2/\lambda = 4\lambda f_\#^2/\pi$ is the Rayleigh range, $r$ the transverse distance, $k = 2\pi/\lambda$ the wavenumber, $\lambda$ the laser wavelength, $w_0$ the laser waist and $f_\#$ the $f$-number. For the two electrons colliding head-on with the laser pulse the relative difference in maximum field strength felt by the two, again not accounting for the laser phase, is to first order given by
\begin{equation}\label{eq:amp}
\frac{\Eback}{\Efront} \approx \sqrt{\frac{16\zR^2+(L+2c\Delta T)^2}{16\zR^2+(L-2c\Delta T)^2}},
\end{equation}
where $\Delta T$ is the time delay between the electrons' centre of mass and the laser pulse at optimal focus. Equation~\ref{eq:amp} is obtained under the assumption that the laser pulse duration is negligible ($c\tau \ll \zR$). In the other extreme, where $c\tau/\zR \rightarrow \infty$, $\Eback/\Efront \rightarrow 1$ as both particles experiences the maximum amplitude at focus and that any envelope tapering becomes imperceptible. For a laser pulse of finite duration the relation between $\zR$, $c\tau$, $c\Delta T$ and $L$ becomes more complicated, but for common parameter values, as presented in Figure~\ref{fig:amp} with $c\tau = \unit[12]{\mu m}$ (\unit[40]{fs}) and $\zR = \unit[1]{\mu m}$ ($f/2$), Equation~\ref{eq:amp} remain a decent estimate. We can further derive an expression for the approximate values of $c\Delta T/\zR$ that maximises Equation~\ref{eq:amp} for a given $L/\zR$,

\begin{equation}\label{eq:opt}
c\Delta T/\zR = \pm\sqrt{(L/2\zR)^2 + 4},
\end{equation}
giving us an expression for the value of this maximum as a function of the distance between the two electrons $L$,
\begin{equation}\label{eq:??}
\max\frac{\Eback}{\Efront} \approx \sqrt{\frac{\sqrt{1 + (4\zR/L)^2} + 1}{\sqrt{1 + (4\zR/L)^2} - 1}}.
\end{equation}
For example, assuming perfect synchronisation between the laser pulse and the leading electron ($c\Delta T = -L/2$) the trailing electron will feel an amplitude of $\Eback = \Efront/\sqrt{2}$ for $L=2\zR\approx\unit[8]{\mu m}$ under the laser and focusing conditions specified above.

\begin{figure*}[t!]
\includegraphics[width=0.99\textwidth]{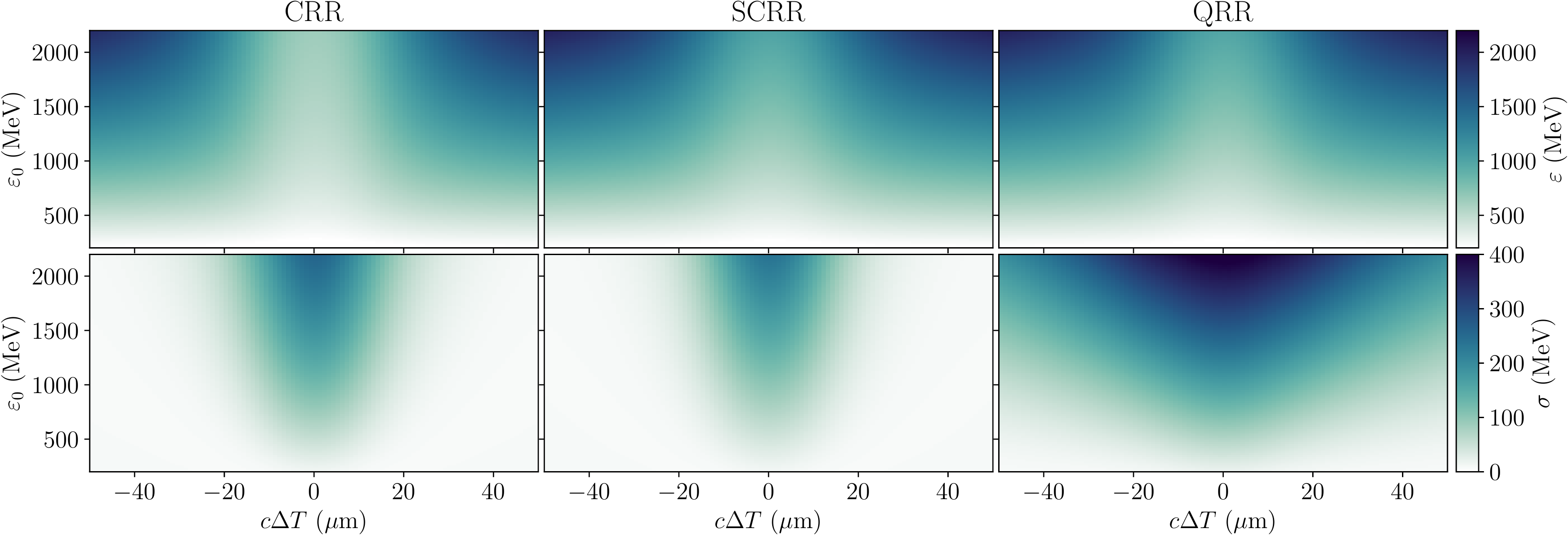}
\caption{\label{fig:output}Mean electron energy $\varepsilon$ (top panels) and standard deviation (bottom panels) after colliding an electron micro-bunch with a tightly focused ($f/2$) laser pulse, presented as functions of pulse delay $c\Delta T$ and initial (mean) electron energy $\varepsilon_0$. The results are shown for three different models of RR; classical (left panels), semi-classical (middle panels) and quantum (right panels).}
\end{figure*}

\section{1D simulations}\label{sec:1d-sims}
To investigate the effect an energy chirp of the electron bunch has on the resulting particle statistics, we perform a large number of single-particle Monte Carlo simulations using the code CIRCE. In order to qualify the effect, we restrict ourselves to varying two primary parameters of interest; the electron initial energy ($\varepsilon_0$) and the temporal delay ($\Delta T$) of the laser pulse. More accurately, we simulate Gaussian-shaped micro-bunches of electrons with both length and width of $\unit[1]{\mu m}$ (FWHM), a mean initial energy of $\varepsilon_0$ and an energy spread $\sigma_0 = \unit[25]{MeV}$ (FWHM). Only head-on collisions are considered and the impact parameter of the bunch is exclusively set to zero.

The spatial dependence of the field is treated as a tightly focused Gaussian beam with waist size $w_0$ and Rayleigh length $z_\mathrm{R} = \pi w_0^2/\lambda$. Going beyond the paraxial approximation, the fields are computed up to fourth-order in the diffraction angle ($w_0/z_\mathrm{R}$), following Ref.~\cite{Salamin.apb.2006}. Although the $f$-number is a principally important parameter that affects the interaction primarily through $\zR$, we have elected to limit ourselves to $f/2$-focusing in order to maintain a reasonable scope of the paper. Similarly, we restrict ourselves to a typical laser wavelength of $\lambda = \unit[0.8]{\mu m}$ and laser pulse duration of $c\tau = \unit[12]{\mu m}$ (\unit[40]{fs}) FWHM. Furthermore, all simulations are carried out for a fixed peak $a_0 = 30$, corresponding to a laser energy of $\unit[2.7]{J}$.

The simulations are generally performed for three different models of radiation reaction: 1) classical radiation reaction (CRR) in the form of the Landau-Lifshitz radiation reaction (LL)~\cite{landau.lifshitz}; 2) semi-classical radiation reaction (SCRR), in the form of LL but with the Gaunt factor correction~\cite{erber.rmp.1966, neitz.prl.2013, niel.pre.2018}; and 3) quantum radiation reaction (QRR), in the form of QED radiation reaction under the LCFA approximation~\cite{dipiazza.prl.2010, ridgers.jcp.2014, gonoskov.pre.2015}.

The results of the simulations are collected in the form of energy spectra and moments, which for each set of parameters in the parameter scan has been averaged over $10^5$ initial electrons. For the electrons we restrict ourselves to the two lowest energy moments, the mean energy $\varepsilon$ and the standard deviation $\sigma$, as they are sufficient to capture most of the spectral information of the micro-bunches. We also capture the generated photon spectra, which will be analysed further in the next section. 

\begin{figure}[b!]
\includegraphics[width=1\columnwidth]{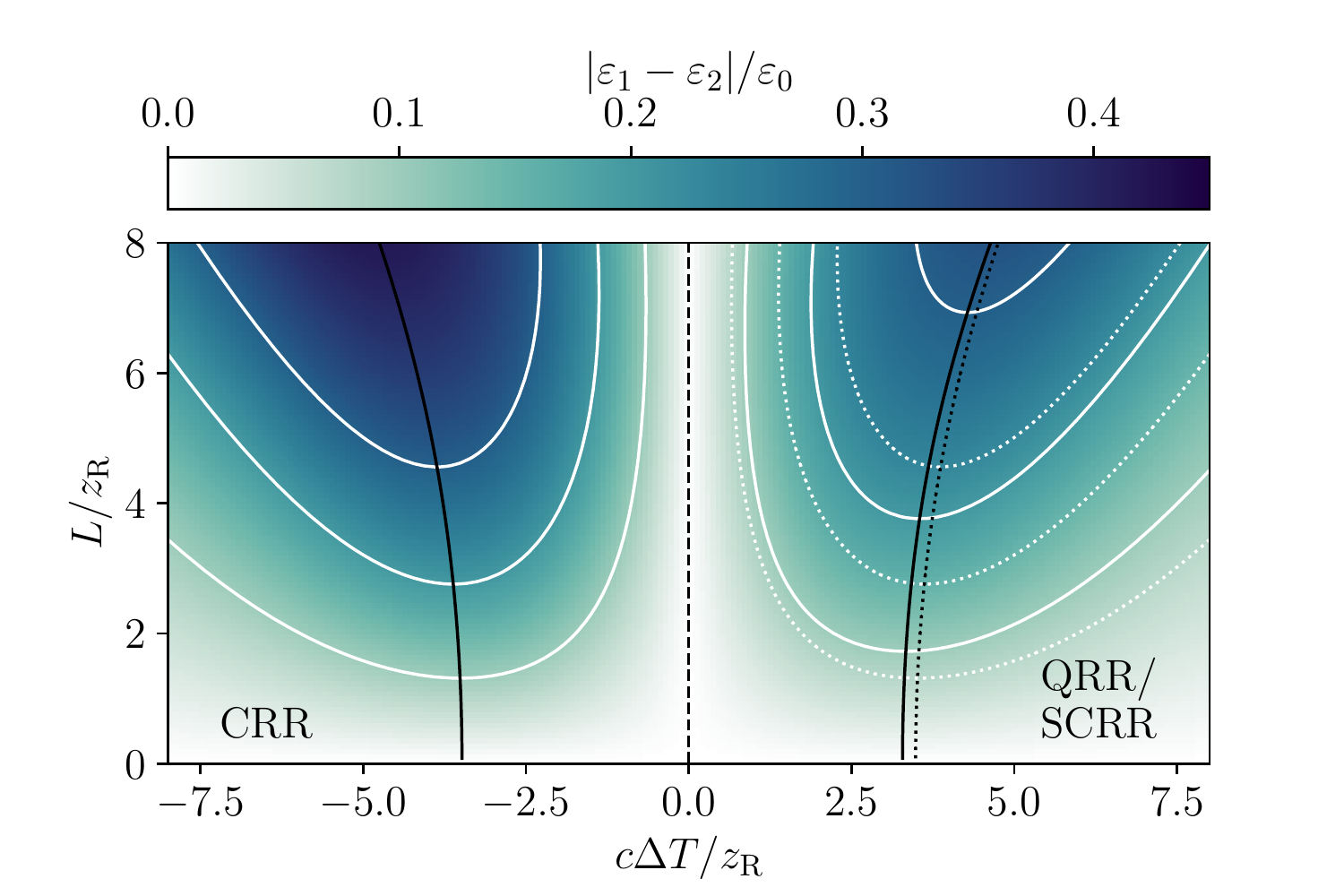}
\caption{\label{fig:mean}Difference in mean energy loss between two electron micro-bunches with initial mean energy of $\varepsilon_0 = \unit[1]{GeV}$, as a function of distance $L$ between the bunches and pulse delay $\Delta T$, and normalised to $\varepsilon_0$. The values are symmetric with respect to $\Delta T$ and so CRR is presented on the left of the dashed line and SCRR/QRR (which are here identical) are shown on the right. Contour lines are show for values of 0.1, 0.2 and 0.3 (white), as well as the position of the maximum for each choice of $L$ (black). Dotted lines on the right show the corresponding CRR values, mirroring the solid lines and contours on the left.}
\end{figure}

\begin{figure*}[t!]
\includegraphics[width=0.67\textwidth]{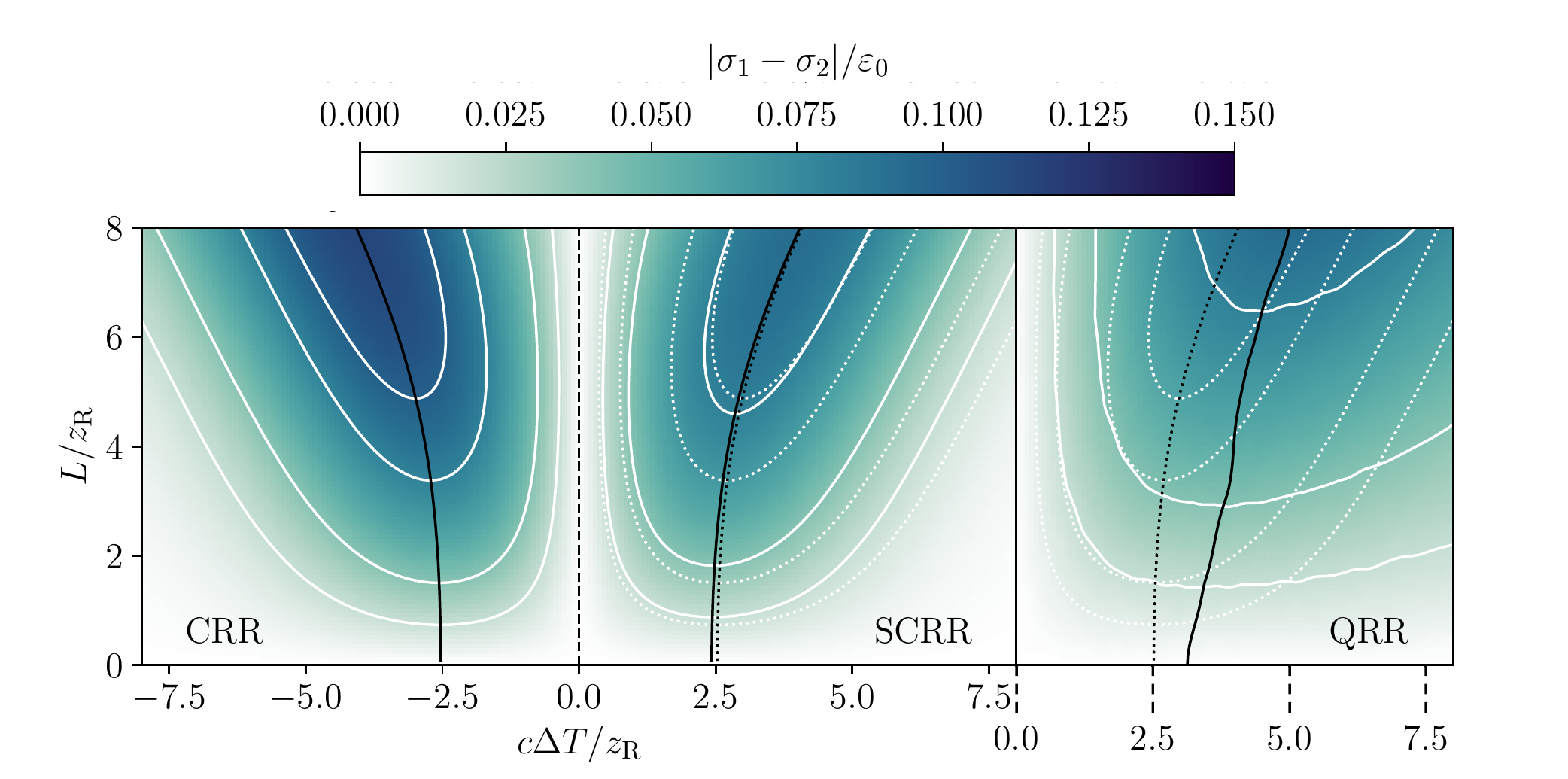}
\caption{\label{fig:std}Difference in energy standard deviation between two electron micro-bunches with initial mean energy of $\varepsilon_0 = \unit[1]{GeV}$, as a function of distance $L$ between the bunches and pulse delay $\Delta T$, and normalised to $\varepsilon_0$. The values are symmetric with respect to $\Delta T$ and so CRR is presented on the left, SCRR in the middle, and QRR is shown on the right. Contour lines are show for values of 0.02, 0.04, 0.08 and 0.10 (white), as well as the position of the maximum for each choice of $L$ (black). Dotted lines in the middle and on the right show the corresponding CRR values, mirroring the solid lines and contours on the left.}
\end{figure*}

The mean electron energy and standard deviation can be seen in Figure~\ref{fig:output} for the three different RR models after the interaction. Here we see that the energy losses are maximised for $c\Delta T = 0$ when the micro-bunch and laser pulse are perfectly synchronised at focus, as is expected. The fact that the losses are symmetric, across positive and negative delays, is a direct consequence of the symmetric laser pulse envelope. Looking closely, the mean energy loss of SCRR and QRR appear identical and slightly lower than that of CRR. However, the main difference between the three models can be seen in the second moment, where QRR is accompanied by a significant stochastic broadening~\cite{neitz.prl.2013, ridgers.jpp.2017} and SCRR by a radiative cooling~\cite{yoffe.njp.2015, vranic.njp.2016}.

With this data, we may now look at the difference in energy loss between two co-propagating micro-bunches longitudinally separated by a distance $L$. The scenario is identical to that described in section~\ref{sec:motivation} and shown in Figure~\ref{fig:setup}. 
For simplicity, both micro-bunches are assumed to have an initial (mean) energy of $\varepsilon_0 = \unit[1]{GeV}$. The relative difference in energy loss for the two micro-bunches is shown in Figure~\ref{fig:mean} as a function of distance $L$ and pulse delay $\Delta T$ for all three RR models. Here we see that the dependence on $L$ and $\Delta T$, unsurprisingly, largely resembles that shown for the difference in maximum pulse amplitude experienced by the two bunches, presented earlier in Figure~\ref{fig:amp}. The main difference between the three models is that CRR predicts a greater difference in mean energy loss between the two bunches, than the semiclassical and quantum models do. In numerical terms, differences of $20$--$30\%$ is common for moderate bunch distances and synchronisation offsets.

The difference in energy spread between the two micro-bunches is similarly presented in Figure~\ref{fig:std} for the three RR models. The classical and semiclassical models show a similar dependence on $L$ and $\Delta T$, but with the SCRR predicting slightly smaller differences between the two bunches. The QRR model, on the other hand, predicts even smaller differences than the SCRR model at moderate synchronisation offsets ($c\Delta T \sim 3\zR$), but this difference extends to much greater pulse offsets ($c\Delta T \gtrsim 6\zR$) where the CRR and SCRR models predicts negligible differences in energy spread between the two bunches. For moderate distances and offsets, the difference in energy spread is typically $5$--$10\%$ relative to the initial energy.

\section{The effect on a finite-sized chirped electron beam}\label{sec:chirped-sims}
Although comparing the effect of different radiation reaction models between two spatially displaced micro-bunches can be instructive, as was done in the previous section, such a setup is not readily available experimentally. In typical high intensity laser-beam interaction experiments, electron beams are generally both longer and contains a broader energy spectrum. Furthermore, we have thus far not fully discussed the role played by the initial energy, apart from that shown in Figure~\ref{fig:output}. While an analysis of how the difference in initial energy between the two micro-bunches affects the interaction outcome could have be made, we have elected to instead consider a more realistic setup.

By combining several micro-bunches, as studied in the previous section, it is possible to effectively construct a finite sized electron beam of both variable length, longitudinal profile and energy spectrum. Moreover, by correlating the micro-bunch position (through $c\Delta T$) with its energy, it also becomes possible to impose a tailored energy chirp, although at the expense of control over the longitudinal profile. For simplicity we here model the electron beam as being Gaussian in shape, both in position as well as energy. More accurately, and in order to reduce the number of confounding factors, we model the beam in position-energy space according to a bivariate Gaussian distribution
\begin{align}
    f(\vec{x}) &= \frac{\exp\!\big( -\frac{1}{2}(\vec{x}-\vec{\mu})^\intercal \mathbf{\Sigma}^{-1} (\vec{x}-\vec{\mu}) \big)}{2\pi\sqrt{|\mathbf{\Sigma}|}},\\
    &\quad\vec{x} = \begin{pmatrix}x\\E\end{pmatrix},\quad
    \vec{\mu} = \begin{pmatrix}\mu_x\\\mu_E\end{pmatrix},\\
    &\quad\mathbf{\Sigma} = \begin{pmatrix}\sigma_x^2 & \rho\sigma_x\sigma_E\\ \rho\sigma_x\sigma_E & \sigma_E^2\end{pmatrix}
\end{align}
where $\mu_x$ ($\mu_E$) is the positional (energy) mean of the electron beam, $\sigma_x$ ($\sigma_E$) the longitudinal (energy) standard deviation, and $\rho$ is a dimensionless chirp parameter on the range $[-1,1]$. The main benefit of this model for the electron beam distribution is that its marginal distributions are entirely independent of $\rho$. Under the assumption that both marginal distributions are either known or constant, this allows us to study the remaining effect due to chirp. With $\rho=0$ energy and position are perfectly uncorrelated while $\rho=\pm 1$ indicates a perfect correlation, signifying a linear chirp.

\begin{figure}[t!]
\includegraphics[width=1\columnwidth]{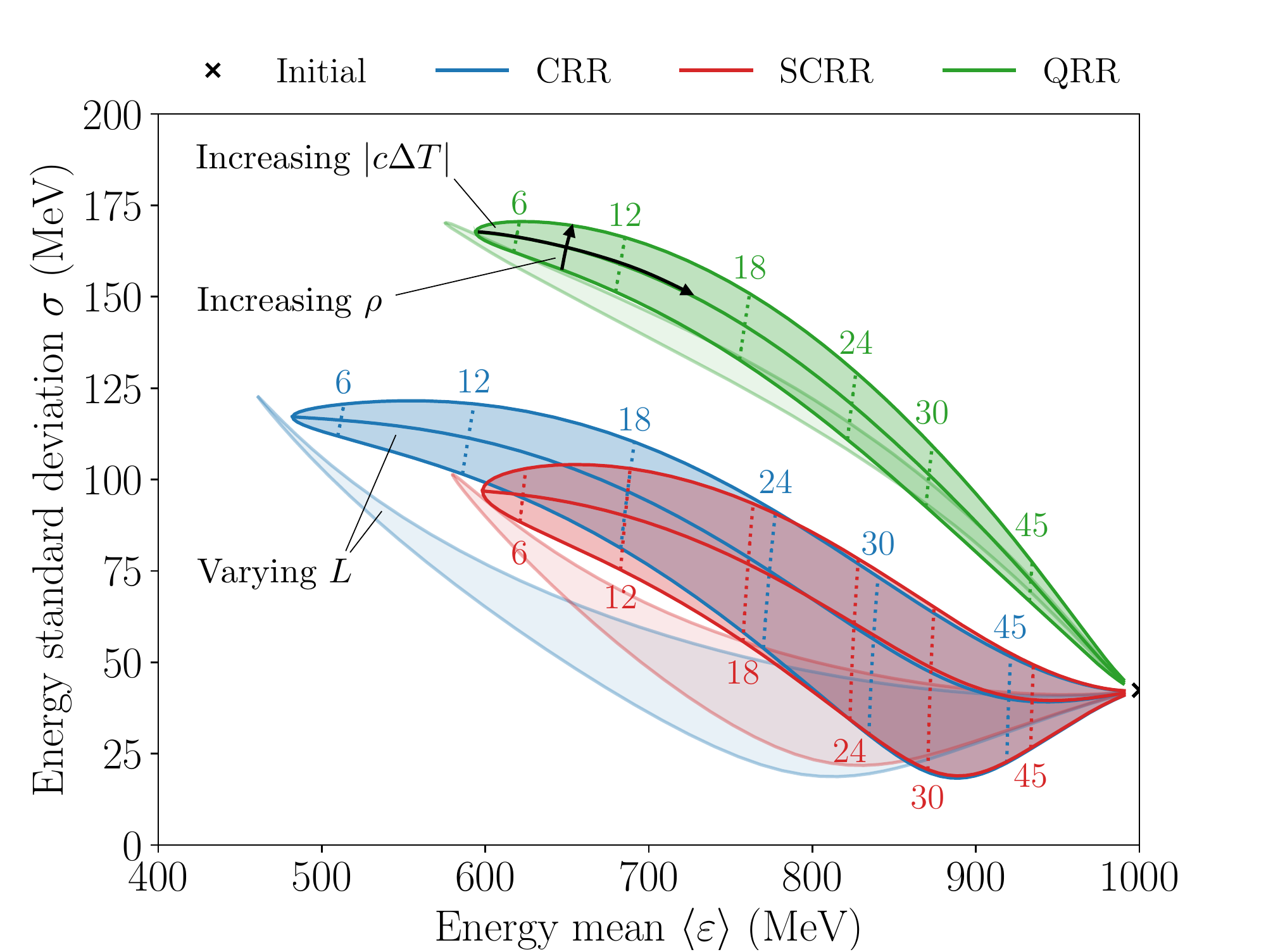}
\caption{\label{fig:mean-std}Contours of the outcome space, in electron energy mean and standard deviation, for a two-dimensional parameter scan across chirp $\rho$ and synchronisation offset $c\Delta T$ for three different radiation reaction models: classical (orange), semiclassical (green) and quantum (red). Two different electron beam lengths are shown: $3\zR$ $(\unit[12]{\mu m})$ (solid contour) and $\zR$ $(\unit[4]{\mu m})$ (transparent contour). Values of constant synchronisation offset are presented in units of micrometers, indicated by the largely vertical dotted contour lines. Similarly, the largely horizontal solid contour lines indicate $\rho=0.9$ (top), $\rho=0$ (middle) and $\rho=-0.9$ (bottom). The electron beam has an initial mean energy of $\varepsilon_0 = \unit[1]{GeV}$ and a spectral width of $\unit[100]{MeV}$ (FWHM) ($\sigma = \unit[42]{MeV}$).}
\end{figure}

We here simulate the interaction with a chirped electron beam and a tightly focused laser pulse, and perform a parameter scan across both chirp $\rho$ and synchronisation offset $\Delta T$. The electron beam is for the entire scan simulated with an initial mean energy of $\varepsilon_0 = \unit[1]{GeV}$ and a spectral width of $\unit[100]{MeV}$ (FWHM). The laser pulse parameters remain unchanged from the previous section. The mean electron energy and standard deviation are computed for each simulation in the parameter scan and the results are presented in Figure~\ref{fig:mean-std} for all three RR models and for two different electron beam lengths $L$ of $\zR$ ($\unit[4]{\mu m}$) and $3\zR$ ($\unit[12]{\mu m}$), both given in FWHM. It should be noted that because of the symmetry of both the laser and electron temporal envelopes the results are symmetric for $c\Delta T \rightarrow -c\Delta T$, $\rho \rightarrow -\rho$. The figure shows that the chirp almost exclusively affects the second energy moment for all three models, and does so about twice as much in the CRR and SCRR models compared to QRR. Furthermore, the CRR and SCRR models have a substantial overlap in the outcome space and are only really separable for very short synchronisation offsets ($|c\Delta T| < \unit[12]{\mu m}$), or for very short electron beam lengths ($L < \unit[4]{\mu m}$). The QRR model is more easily distinguishable from the other two, but primarily through the second moment as it is identical to the SCRR model in the first moment.

\begin{figure}[t!]
\includegraphics[width=1\columnwidth]{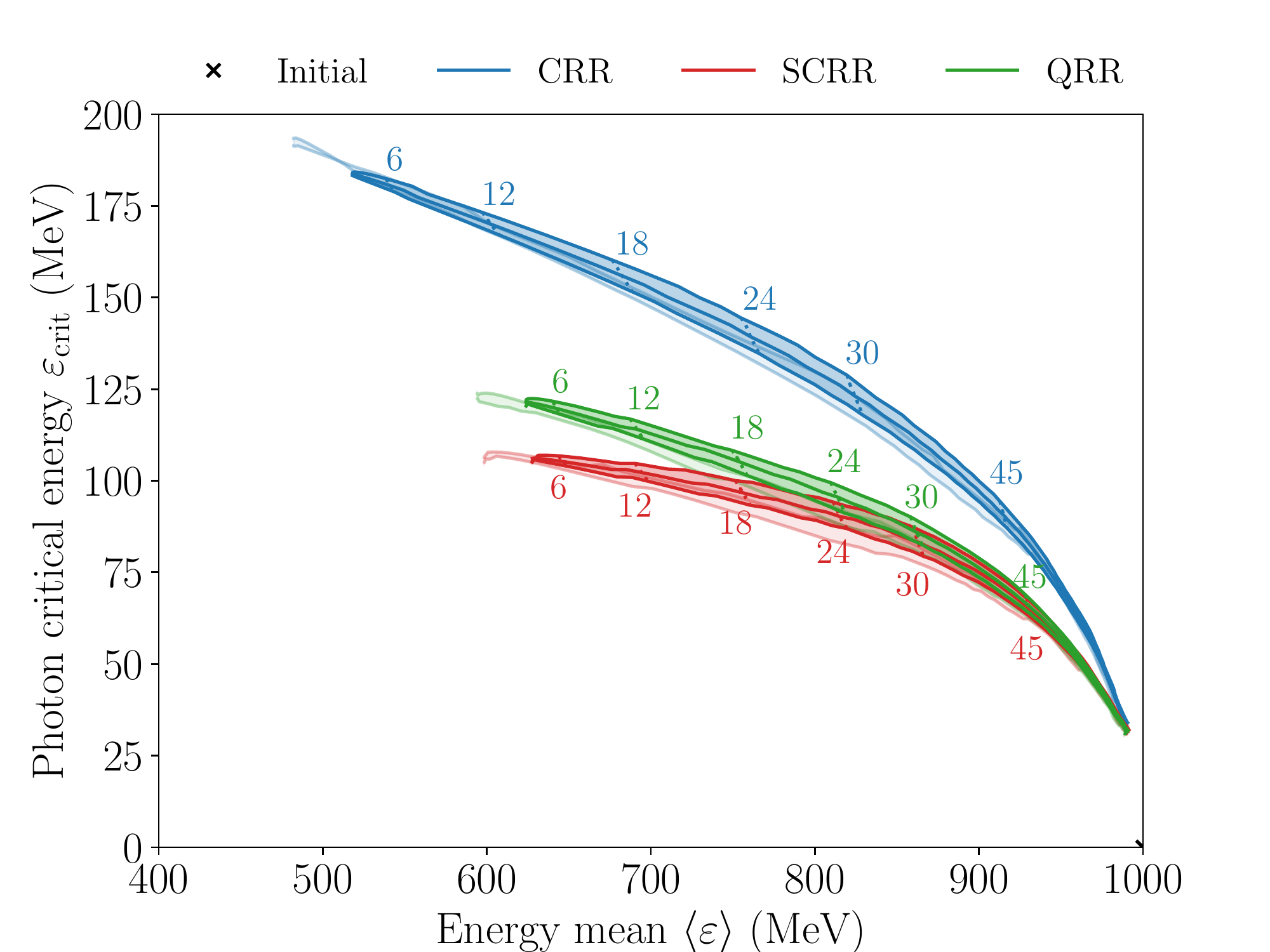}
\caption{\label{fig:mean-photon}Conditions as in Figure~\ref{fig:mean-std}, but here presented in terms of mean electron energy and photon critical energy. Two different electron beam lengths are shown: $5\zR$ $(\unit[20]{\mu m})$ (solid contour) and $3\zR$ $(\unit[12]{\mu m})$ (transparent contour).}
\end{figure}

We also analyse the photon spectrum generated in the interaction. In order to condense it down to a singular number, the photon power spectra are fitted to the functional form $x^{1/3}\e^{-(x/\varepsilon_\mathrm{crit})^C}$, where $\varepsilon_\mathrm{crit}$ is the photon critical energy and $C$ is a fitting parameter\footnote{$C$ is fitted for each spectrum separately.} used to compensate for the fact that the obtained photon spectra are broader than that of a mono-energetic electron beam. The results are presented in Figure~\ref{fig:mean-photon}, as before for all three models, but now in terms of the mean electron energy and photon critical energy and for two electron beam lengths $L$ of $3\zR$ ($\unit[12]{\mu m}$), and $5\zR$ ($\unit[20]{\mu m}$) (FWHM). Here it can be seen that the chirp has a much smaller effect on the photon critical energy than it did on the electron second energy moment. This is also the motivation for the choice of longer beam lengths, in order to make the variation more visible. Similarly to before, we also see that it is primarily for larger synchronisation offsets that the difference between different RR models become indistinguishable. However, unlike for the second moment, the photon critical energy is mainly comparable between the SCRR and QRR models.

\begin{figure}[t!]
\includegraphics[width=1\columnwidth]{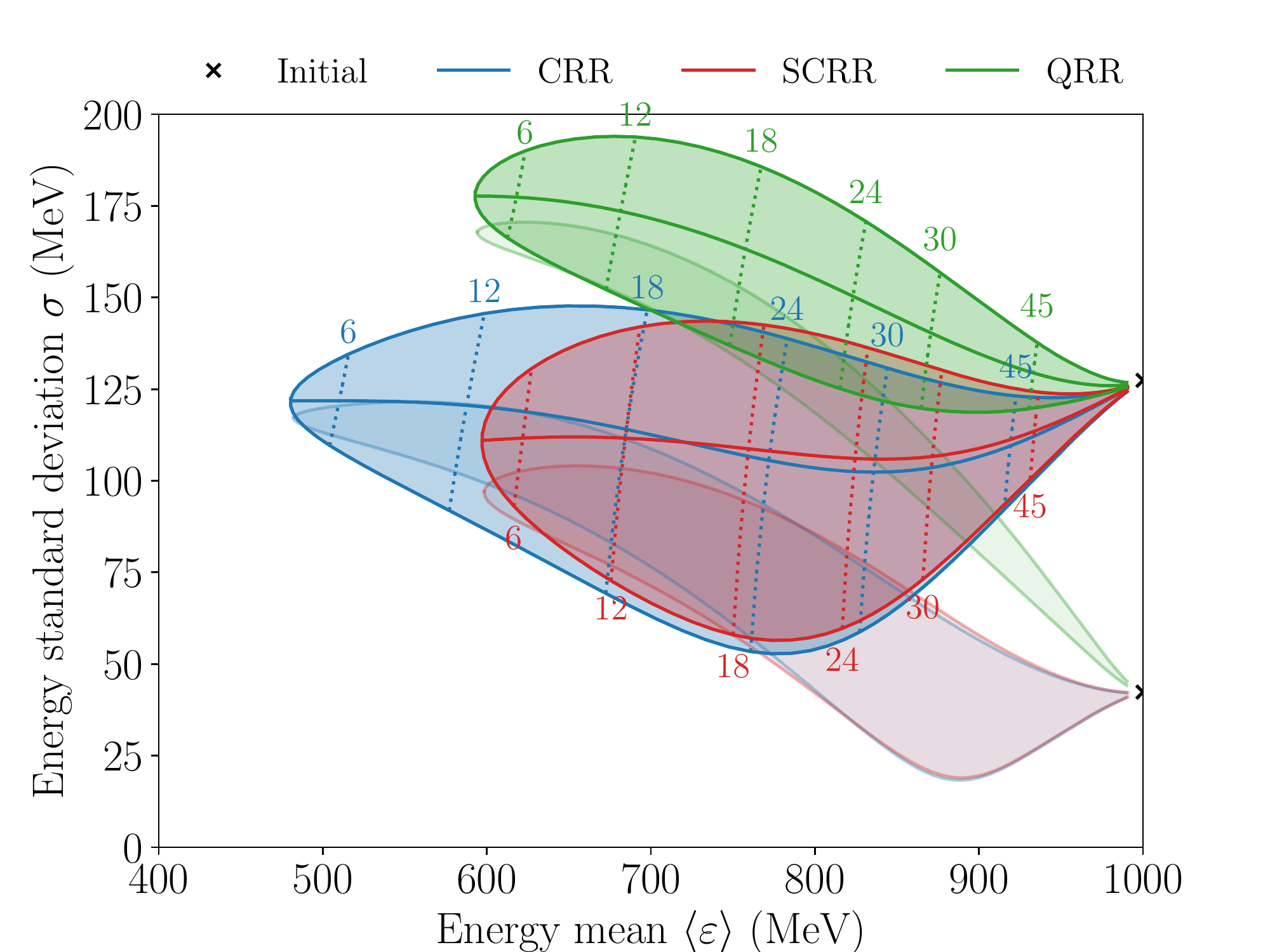}
\caption{\label{fig:mean-std2}Conditions as in Figure~\ref{fig:mean-std}, but here presented for an electron beam length of $3\zR = \unit[12]{\mu m}$ and an initial spectral width of $\unit[300]{MeV}$ (FWHM, solid contour) ($\sigma = \unit[127]{MeV}$). The results with an initial spectral width of $\unit[100]{MeV}$ (FWHM) ($\sigma = \unit[42]{MeV}$) is shown for reference (transparent contour).}
\end{figure}

Finally, in Figure~\ref{fig:mean-std2} we present how a change in the initial spectral width to $\unit[300]{MeV}$ (FWHM) affects the outcome. Unsurprisingly, a broader initial energy spectrum enhances the effect of the chirp and primarily in the second energy moment, such that the QRR model now partially overlap the outcome space of the CRR and SCRR models. For synchronisation offsets $c\Delta T$ greater than about $\unit[20]{\mu m}$ it may thus become impossible to distinguish, in the first and second electron energy moments alone, between the three models on a shot-to-shot basis, unless more information is known about the chirp.

\section{Conclusions}
We have investigated the role of energy chirp in the interaction of a high-energy electron beam and a tightly focused laser pulse, identified which parameters are the most significant and demonstrated the variation across different radiation reaction models, of relevance for current and future experimental campaigns. We generally find that the effect of the energy chirp is predominantly expressed through the second energy moment. We have also shown that the strength of the effect is dependent on the synchronisation between the laser pulse and the electron beam, as well as the length and spectral width of the electron beam.

For certain parameter choices, e.g. when the synchronisation between the laser pulse and electron beam is not perfect, the outcome space of the different radiation reaction models begin to overlap. The semiclassical and quantum models are near identical in first moment, but shows significant separation in the second moment
up to moderate synchronisation offsets, so long as the spectral width of the electron beam is not too large. The two models also produce some separation in photon critical energy, but again show significant overlap for greater synchronisation offsets. The classical radiation reaction model on the other hand, distinguishes itself from the other two in both the first energy moment and photon critical energy, although more so through the latter. For greater synchronisation offsets the classical model becomes practically indistinguishable from the semiclassical model in all but the photon spectrum.

Taken together, the effect of the chirp is not so large as to inhibit the distinguishability of the three models, but sufficiently large to have an impact, in particular on a shot-to-shot basis and more so the larger the synchronisation offset. However, these issues can in general be overcome by gathering statistics across multiple shots or through more detailed knowledge of the energy chirp, such that the parameter space can be constrained. Furthermore, the photon spectrum can be used as an important discriminator as it is largely unaffected by the chirp.

\begin{acknowledgments}
The research is supported by the Swedish Research Council grants No. 2013-4248, 2016-03329, 2020-06768 (M.M.), the Knut \& Alice Wallenberg Foundation (J.M., M.M.), the Engineering and Physical Sciences Research Council UK grants No. EP/V049461/1 (C.P.R.),  EP/V049186/1 (E.G.) and EP/V049577/1 (S.P.D.M), and by Horizon 2020 funding under European Research Council (ERC) Grant Agreement No. 682399. 
The simulations were performed on resources provided by the Swedish National Infrastructure for Computing (SNIC) at HPC2N.
\end{acknowledgments}

\bibliography{references}{}
\bibliographystyle{aipnum4-1}

\vfill

\end{document}